\documentclass[aps,onecolumn]{revtex4}
\begin{document}
\draft
\author{Yi-shi Duan and Peng-ming Zhang\thanks{%
Corresponding author. Email: zhpm@lzu.edu.cn}}
\title{Inner structure of Gauss-Bonnet-Chern Theorem and the Morse theory}
\date{May,4 2001 }
\address{Institute of Theoretical Physics, Lanzhou University, Lanzhou, 730000,
People' Republic of China}
%\maketitle

\begin{abstract}
We define a new 1-form $H^A$ based on the second fundamental tensor $H_{ab%
\overline{A}}$, the Gauss-Bonnet-Chern form can be novelly expressed with
this 1-form. Using the $\phi $-mapping theory we find that the
Gauss-Bonnet-Chern density can be expressed in terms of the $\delta $%
-function $\delta (\vec{\phi})$ and the relationship between the
Gauss-Bonnet-Chern theorem and Hopf-Poincar\'{e} theorem is given
straightforward. The topological current of the Gauss-Bonnet-Chern theorem
and its topological structure are discussed in details. At last, the Morse
theory formula of the Euler characteristic is generalized.
\end{abstract}


\maketitle
\section{Introduction}

The Gauss-Bonnet-Chern (GBC) theorem is one of the most significant results
in topology, which relates the curvature tensor of a compact and oriented
even-dimensional Riemannian manifold $M$ to an important topological
invariant, the Euler characteristic $\chi (M)$ of $M$. A great advance in
this field is the discovery of the relationship between supersymmetry and
the index theorem, which includes the derivation of the GBC theorem via
supersymmetry and the path integral techniques as presented by
Alvarez-Gaum\'{e} et al. \cite{Alvarez}. On the physics side, the optical
Berry phase is a direct result of the Gauss-Bonnet theorem \cite{Berry} and
the black hole entropy emerges as the Euler class through dimensional
continuation of the Gauss-Bonnet theorem \cite{entropy}. Furthermore, it
must be pointed out that the topological current of GBC theorem is important
to study the topological defects in Ginzburg-Landau theory et al. \cite
{Mazenko,Ying}

It is well-known that there were three theoretical ways to study the Euler
characteristic $\chi (M)$ of a compact oriented even-dimensional Riemannian
manifold $M$. The first is the integral of $GBC$ density in terms of Chern
form \cite{Chern1}; the second is the Hopf-Poincar\'{e} theorem \cite{Hopf}
which shown that the $\chi (M)$ is the sum of the indices of the zeroes of
the vector field on $M$; and the third is the so called Morse theory \cite
{Milnor2} which related to the Hessian Matrices. The main purpose in this
paper is to present a new theoretical framework which directly gives the
inner relationships between $GBC$ theorem, Hopf-Poincar\'{e} theorem and the
Morse theory. For this purpose we introduce a new 1-form $H^A$ based on the
second fundamental tensor $H_{ab},$ and give a decomposed expression of the
curvature 2-form $F^{AB}$. Using this expression we find an important
formula of the GBC form, which is simply determined by the edge production
of the 1-form $H^A$. It gives a natural way to obtain the $\delta $-function
form of the GBC density with the $\phi $-mapping theory \cite{Slac,Duan}. By
making use of the above formulation the GBC theorem is proved to be identify
with the Hopf-Poincar\'{e} theorem. It is noted that the Chern's form and
the gauge potential decomposition of SO(n) \cite{MengLi} are unnecessary in
the proof. Furthermore, according to the topological structure of the GBC
density, a generalization of Morse theory can be obtained directly when the
vector field on the manifold is considered as the gradient field of a Morse
function $f$.

In Section II we give the definition of the 1-form $H^A$ and the expression
of the curvature 2-form $F^{AB}$ in the Riemannian manifold. In Section III,
in terms of the $\phi $-mapping theory we get the $\delta $-function
expression of the $GBC$ density and study the topological structure of the $%
GBC$ density, in which the Hopf index and Brouwer degree of $\phi $-mapping
at the zero points of $\phi (\vec{x})$ play an important role. The
topological current of $GBC$ theorem is given and the general velocity of
the $i$-th zero of the vector field $\phi (\vec{x})$ is studied in detail in
Section IV. At last, we discuss the relationship between Euler
characteristic $\chi (M)$ and indices of the critical points of function $f$
and get the generalized formula of $\chi (M)$ in the Morse theory in Section
V.

\section{A new 1-form $H^A$ in Riemannian manifold}

Let $M$ be a compact and oriented $n$(even)-dimensional Riemannian manifold,
which can be immersed in a Euclidean space $R^{m+n}(m+n=n(n+1)/2)$ with
local coordinate $x^\mu $ which satisfy the parametric equation \cite
{Eisenhart}
\[
x^\mu =x^\mu (u^1,u^2,...,u^n),\;\;\;\;\;\;\mu =1,2,...,m+n
\]
where $u^a$ is the local coordinate of $M$. In the $(m+n)$-dimensional
Euclidean space, the tangent vector $B_a^\mu $ on the $M$ is defined as
\[
B_a^\mu =\frac{\partial x^\mu }{\partial u^a},\;\;\;\;\;a=1,2,...,n
\]
and the metric tensor $g_{ab}$ in the $M$ is determined by
\[
g_{ab}=B_a^\mu B_b^\mu .
\]
The unit normal vector $N_{\overline{A}}^\mu $ satisfy
\begin{equation}
N_{\overline{A}}^\mu B_a^\mu =0,\;\;\;\;\;\overline{A}=n+1,n+2,...,(m+n),
\label{nor1}
\end{equation}
and the second fundamental tensor $H_{ab\overline{A}}$ is introduced as
follow:
\begin{equation}
H_{ab\overline{A}}=N_{\overline{A}}^\mu \nabla _aB_b^\mu ,\;\;\;\;\mu
=1,2,...,(m+n),\text{\ \ \ \ \ }a,b=1,2,...,n.  \label{h1}
\end{equation}

Using the Gauss-Codazzi formula \cite{Eisenhart} the curvature tensor $%
R_{ab,cd}$ is expressed with the second fundamental tensor
\[
R_{ab,cd}=H_{ac\overline{A}}H_{bd\overline{A}}-H_{ad\overline{A}}H_{bc%
\overline{A}}.
\]
Then the curvature tensor of the SO(n) principle bundle, i.e., the SO(n)
gauge field tensor $F_{ab}^{AB}=-R_{ab,cd}e^{Ac}e^{Bd}$ can be written as
\[
F_{ab}^{AB}=(H_{ac\overline{A}}H_{bd\overline{A}}-H_{ad\overline{A}}H_{bc%
\overline{A}})e^{Ac}e^{Bd},
\]
where $e^{Aa}$ is the veilbein on $M$, which is defined by
\[
g^{ab}=e^{Aa}e^{Ab},\;\;\;\;\;A=1,2,...,n.
\]
It is well known that the metric $g^{ab}$ is invariant under the SO(n) gauge
transformation
\[
e^{^{\prime }Aa}=L_B^A(u)e^{Ba},\;\;\;\;\;\;\;\;L_B^A(u)\in SO(n).
\]
From the above formulation we see that if we define a 1-form $H_{\overline{A}%
}^A$%
\begin{equation}  \label{ha1}
H_{\overline{A}}^A=e^{Aa}H_{ab\overline{A}}du^b,
\end{equation}
the field strength 2-form $F^{AB}$ can be expressed as
\begin{equation}  \label{fa1}
\begin{array}{c}
F^{AB}= \frac 12F_{ab}^{AB}du^a\wedge du^b \\
=H_{\overline{A}}^A\wedge H_{\overline{A}}^B.
\end{array}
\end{equation}

\section{Gauss-Bonnet-Chern density}

The GBC form $\Lambda $ is a $n$-form over $M,$ i.e.
\begin{equation}  \label{h6}
\Lambda =\frac{(-1)^{n/2}}{2^n\pi ^{n/2}(n/2)!}\varepsilon
^{A_1A_2...A_n}F^{A_1A_2}\wedge F^{A_3A_4}\wedge ...\wedge F^{A_{n-1}A_n}.
\end{equation}
The famous GBC theorem can thus be expressed as
\begin{equation}  \label{euler}
\chi (M)=\int_M\Lambda .
\end{equation}
With the 1-form $H_{\overline{A}}^A$ we can obtain a new elegant expression
of the GBC form
\begin{equation}  \label{g1}
\Lambda =\frac{(-1)^{n/2}}{2^n\pi ^{n/2}(n/2)!}\varepsilon ^{A_1A_2...A_n}H_{%
\overline{A}_1}^{A_1}\wedge H_{\overline{A}_1}^{A_2}\wedge ...\wedge H_{%
\overline{A}_{n/2}}^{A_n}.
\end{equation}

Let $n^a(a=1,2,...,n)$ be a unit tangent vector on $M$
\[
n^an^a=1,
\]
by which two new unit tangent vectors $n^\mu $ and $n^A$ can be defined as
\begin{equation}  \label{h2}
n^\mu =B_a^\mu n^a,\;\;\;n^A=e^{Aa}n_a,
\end{equation}
and it is easy to verify that
\begin{equation}  \label{cin}
n^\mu N_{\overline{A}}^\mu =0,\;\;\;\;\;n^An^A=1.
\end{equation}
From Eqs. (\ref{h1}) and (\ref{ha1}) one can find
\begin{equation}  \label{per1}
H_{\overline{A}}^An^A=N_{\overline{A}}^\mu dn^\mu .
\end{equation}
To study the topology of manifold $M$ by making use of the unit vector field
$n^\mu $ in $R^{(m+n)},$ it must be required that both $n^\mu $ and $dn^\mu $
should be intrinsic on $M$, i.e.,
\begin{equation}  \label{imp1}
N_{\overline{A}}^\mu dn^\mu =0.
\end{equation}
Thus one can find a relation
\begin{equation}  \label{hn}
H_{\overline{A}}^An^A=0.
\end{equation}
It can be seen from Eq. (\ref{ha1}) that the 1-form $H_{\overline{A}}^A$ is
invariant in general covariant coordinates transformation on $M$ and
covariant in the SO(n) gauge transformation, i.e.,
\[
n^{^{\prime }A}=L_B^A(u)n^B,\;\;\;\;H_{\overline{A}}^{^{\prime
}A}=L_B^A(u)H_{\overline{A}}^B.
\]

Since the covariant 1-form $H_{\overline{A}}^A$ is perpendicular to $n^A$,
for each fixed index $\overline{A},$ the only covariant 1-form of $H_{%
\overline{A}}^A$ related to $n^A$ should be
\begin{equation}  \label{h3}
H_{\overline{A}}^A=k_{\overline{A}}(u)Dn^A,
\end{equation}
where $k_{\overline{A}}(u)$ is a scalar function for each fixed index on $M$
and
\[
Dn^A=dn^A-\omega ^{AB}n^B,
\]
in which $\omega ^{AB}$ is the spin connection, i.e., the connection of the $%
SO(n)$ principle bundle. Then from Eq. (\ref{fa1}) the field strength $%
F^{AB} $ can be given
\begin{equation}  \label{h5}
F^{AB}=k^2(u)Dn^A\wedge Dn^B,
\end{equation}
where $k^2(u)=k_{\overline{A}}(u)k_{\overline{A}}(u).$

With the Eqs. (\ref{h6}) and (\ref{h5}) the GBC form can be expressed as
\begin{equation}  \label{chen}
\Lambda =\frac{(-1)^{n/2}k^n(u)}{2^n\pi ^{n/2}(n/2)!}\epsilon
^{A_1A_2...A_n}Dn^{A_1}\wedge Dn^{A_2}\wedge ...\wedge Dn^{A_n}.
\end{equation}
Here following Chern's work \cite{Chern1} and let $P$ be an arbitrary but
fixed point of $R^n$. In the neighborhood of $P$ one can choose a family of
veilbein \{$e^{Aa}$\} such that at $P$
\begin{equation}  \label{ds0}
\omega ^{AB}=0,
\end{equation}
which gives
\begin{equation}  \label{add3}
\Lambda =\frac{(-1)^{n/2}k^n(u)}{2^n\pi ^{n/2}(n/2)!}\epsilon
^{A_1A_2...A_n}dn^{A_1}\wedge dn^{A_2}\wedge ...\wedge dn^{A_n}.
\end{equation}
Since the integral region is the manifold $M$, at the same time the integral
kernel is only function defined in this manifold $M,$ the last result should
be independent of the choice of Euclidean space's dimension. With the help
of (n+1)-dimensional Euclidean space $R^{n+1}$, we can give
\begin{equation}  \label{nwe1}
k=(\frac{n!!}{(n-1)!!})^{1/n},
\end{equation}
which is discussed in detail in the appendix. Then the above GBC form can be
simply expressed as
\begin{equation}  \label{add1}
\Lambda =\frac 1{A(S^{n-1})(n-1)!}\epsilon ^{A_1A_2...A_n}dn^{A_1}\wedge
dn^{A_2}\wedge ...\wedge dn^{A_n},
\end{equation}
which is just the result obtained verbosely by Chern in Ref. \cite{Chern1}.
In the $\phi $-mapping theory \cite{Slac,Duan} the unit vector $n^A$ should
be further determined by the smooth vectors $\phi ^A$, i.e.
\begin{equation}  \label{add2}
n^A=\frac{\phi ^A}{||\phi ||},\;\;\;\;||\phi ||^2=\sqrt{\phi ^A\phi ^A}.
\end{equation}
In fact $n$ is identified as a section of the sphere bundle over $M$ (or a
partial section of the vector bundle over $M$). We see that the zeroes of $%
\phi $ are just the singular points of $n$. Since the global property of a
manifold has close relation with zeroes of a smooth vector fields on it, the
above expression of the unit vector $n$ is a very powerful tool in the
discussion of the global topology.

Substituting Eq. (\ref{add2}) into Eq. (\ref{add1}), the GBC form can be
given as \cite{Slac,MengLi}
\begin{equation}  \label{GBC2}
\Lambda =\delta (\vec \phi )D(\frac \phi u)d^nu
\end{equation}
where the Jacobian $D(\frac \phi u)$ is defined as
\[
\varepsilon ^{A_1A_2...A_n}D(\frac \phi u)=\varepsilon
^{a_1a_2...a_n}\partial _{a_1}\phi ^{A_1}\partial _{a_2}\phi
^{A_2}...\partial _{a_n}\phi ^{A_n}.
\]
We define the GBC density $\rho $ on $M$ as
\begin{equation}  \label{den1}
\rho :=\frac{\varepsilon _{A_1A_2...A_n}\varepsilon ^{a_1a_2...a_n}}{%
A(S^{n-1})(n-1)!}\partial _{a_1}\phi ^{A_1}\partial _{a_2}\phi
^{A_2}...\partial _{a_n}\phi ^{A_n}=\delta (\vec \phi )D(\frac \phi u),
\end{equation}
which shows that only at the zero points of the vector field $\vec \phi (u)$%
, i.e.
\begin{equation}  \label{solution}
\begin{array}{c}
\phi ^1(u^1,u^2,...,u^n)=0, \\
\phi ^2(u^1,u^2,...,u^n)=0, \\
... \\
\phi ^n(u^1,u^2,...,u^n)=0,
\end{array}
\end{equation}
we can get the nontrial GBC density. The expressions (\ref{GBC2}) and (\ref
{den1}) are of great importance: they yield, in our case, the evident result
of the Hopf theorem.

Suppose that the vector field $\vec{\phi}(u)$ possesses $l$ isolated zeroes,
according to the implicit function theorem \cite{Imp1,Imp2}, when the
Jacobian $D(\phi /u)\neq 0$, the solutions of Eq. (\ref{solution}) are
generally expressed as
\[
u_i^A=z_i^A,\;\;\;A=1,2,...,n,\;\;i=1,2,...,l.
\]
In terms of the $\phi $-mapping theory \cite{Slac,Duan} and the $\delta $%
-function theory \cite{delta}, one can rigorously prove that the $\delta (%
\vec{\phi}(u))$ can be expanded as
\[
\delta (\vec{\phi})=\sum_{i=1}^l\frac{W_i\delta (\vec{u}-\vec{z}_i)}{D(\phi
/u)|_{u=z_i}},\;\;\;\;W_i=\beta _i\eta _i,
\]
where $W_i$ is the winding number of the vector field, $\beta _i=|W_i|$ is
the Hopf index and $\eta _i$ is the Brouwer degree of map $x\rightarrow \phi
$ \cite{Hopf}
\[
\eta _i=sgnD(\frac \phi u)|_{u=z_i}=\pm 1.
\]
This leads to the following topological structure
\[
\Lambda =\delta (\vec{\phi})D(\frac \phi u)d^nu=\beta _i\eta _i\delta (\vec{u%
}-\vec{z}_i)d^nu,
\]
which means that GBC form is labeled by the Brouwer degree and Hopf index.
Therefore, the Euler characteristic $\chi (M)$ can be represented as
\begin{equation}  \label{char2}
\chi (M)=\int_M\Lambda =\sum_{i=1}^lW_i.
\end{equation}
Here, Eq. (\ref{char2}) states that the sum of indices of the zeroes of
vector $\vec{\phi}$ is the Euler characteristic. Therefore, the topological
structure of GBC form reveals the expected result of the Hopf-Poincar\'{e}
theorem.

\section{The GBC topological current}

Let us consider the ($n+1$)-dimensional manifold ${\bf M}\times {\bf R}$
with coordinate $u^0=t\in {\bf R}$ denoted as a time variable. The line
element of ${\bf M}\times {\bf R}$ is
\[
ds^2=\bar{g}_{\alpha \beta }du^\alpha du^\beta =(du^0)^2-g_{ab}du^adu^b\quad
(\alpha ,\beta =0,1,\cdots ,n;\;\,\,\;a,b=1,2,\cdots ,n),
\]
which implies
\[
\bar{g}=det(\bar{g}_{\alpha \beta })=-g,\quad \quad \sqrt{-\bar{g}}=\sqrt{g}%
.
\]

A generally covariant GBC topological current is defined as
\[
j^\alpha :=\frac{(-1)^{n/2}}{2^n\pi ^{n/2}(n/2)!}\varepsilon _{A_1A_2...A_n}%
\frac{\varepsilon ^{\alpha \alpha _1\alpha _2...\alpha _n}}{\sqrt{g}}%
F_{\alpha _1\alpha _2}^{A_1A_2}F_{\alpha _3\alpha _4}^{A_3A_4}...F_{\alpha
_{n-1}\alpha _n}^{A_{n-1}A_n}.
\]
It follows that
\begin{equation}  \label{current}
j^\alpha :=\frac 1{A(s^{n-1})(n-1)!}\epsilon _{A_1A_2...A_n}\frac{\epsilon
^{\alpha \alpha _1\alpha _2\cdots \alpha _n}}{\sqrt{g}}\partial _{\alpha
_1}n^{A_1}\partial _{\alpha _2}n^{A_2}\cdots \partial _{\alpha _n}n^{A_n},
\end{equation}
where $j^0$ is just the GBC density $\frac 1{\sqrt{g}}\rho $. Obviously the
current (\ref{current}) is identically conserved,
\[
\bigtriangledown _\alpha j^\alpha =\frac 1{\sqrt{g}}\partial _\alpha (\sqrt{g%
}j^\alpha )=0.
\]
If we define $n+1$ Jacobians as
\begin{equation}  \label{highjacobi}
\epsilon ^{A_1\cdots A_n}D^\alpha (\phi /u)=\epsilon ^{\alpha \alpha
_1\cdots \alpha _n}\partial _{\alpha _1}\phi ^{A_1}\cdots \partial _{\alpha
_n}\phi ^{A_n},
\end{equation}
in which $D^0(\frac \phi u)=D(\frac \phi u)$ is the usual $n$-dimensional
Jacobian. Making use of the Laplacian relation, we obtain the $\delta $%
-function-like current
\begin{equation}  \label{deltacurrent}
j^\alpha =\frac 1{\sqrt{g}}\delta (\vec{\phi})D^\alpha (\phi /u).
\end{equation}
Suppose $\phi ^A(u)$ possesses $l$ isolated zeroes on ${\bf M}$ and let the $%
i$-th zero be $\vec{u}=\vec{z}_i$ i.e.
\begin{equation}  \label{zeropoint}
\phi ^A(\vec{z}_i,t)=0,\quad \quad i=1,2,\cdots ,l,\,\;\;\;\,A=1,2,\cdots ,n,
\end{equation}
according to the implicit function theorem whose solution can be expressed
as \cite{MengLi}
\begin{equation}  \label{trajectory}
\vec{z}_i=\vec{z}_i(t),
\end{equation}
which is the trajectory of the $i$-th zero. One can prove that the general
velocity of the $i$-th zero \cite{Slac,MengLi},
\begin{equation}  \label{velocity}
V^\alpha :=\frac{dz_i^\alpha }{dt}=\frac{D^\alpha (\phi /u)}{D(\phi /u)}|_{%
\vec{u}=\vec{z}_i}.\quad \quad V^0=1.
\end{equation}
Then the topological current $j^\alpha $ can be written as the current
density form of a system including $l$ classical point particles with
topological charge $W_i=\beta _i\eta _i$ moving in the ($n+1$)-dimensional
space-time
\begin{equation}  \label{f564}
j^\alpha ={\frac 1{\sqrt{g}}}\sum\limits_{i=1}^lW_i\delta (\vec{u}-\vec{z}%
_i(t))\frac{dz_i^\alpha }{dt},\qquad j^0={\frac 1{\sqrt{g}}}%
\sum\limits_{i=1}^lW_i\delta (\vec{u}-\vec{z}_i(t))
\end{equation}
The total charge of the system is
\begin{equation}  \label{f565}
W:=\int_{{\bf M}}j^0\sqrt{g}d^nu=\sum\limits_{i=1}^l\beta _i\eta
_i=\sum\limits_{i=1}^lW_i,
\end{equation}
which is none other than the topological invariant $\chi ({\bf M})$, i.e. $%
\chi ({\bf M})=W$. From (\ref{den1}) and (\ref{deltacurrent}), we get a
concise expression for the topological current,
\begin{equation}  \label{f567}
j^\alpha =\rho \frac{D^\alpha (\phi /u)}{\sqrt{g}D(\phi /u)}=\frac 1{\sqrt{g}%
}\rho V^\alpha ,
\end{equation}
which takes the same form as the current density in hydrodynamics. From (\ref
{f565}) the topological charge can be expressed as
\begin{equation}  \label{charge}
W=\int_{{\bf M}}\rho d^nu=\int_{{\bf M}}\delta (\vec{\phi})D({\frac \phi u}%
)d^nu=deg\ n\int_{n({\bf M})}\delta (\vec{\phi})d^n\phi ,
\end{equation}
where $deg\ n$ is the degree of the mapping $n$ \cite{Milnor}. It indicates $%
W=deg\ n$. Expressions (\ref{f564})-(\ref{charge}) show that the topological
structure of the $GBC$ topological current is characterized by the Brouwer
degrees and Hopf indices. The structure of the $GBC$ topological current and
its formulation is of great use in studying the topological defects in the
Ginzburg-Landau theory \cite{Mazenko,Ying} and topological field theories,
especially in low-dimensional cases \cite{Duan}.

In our theory the point-like particles with topological charges $W_i=\beta
_i\eta _i(i=1,2,\cdots ,l)$ are called $GBC$ topological particles. These
particles are just located at the zeros of $\vec \phi (u)$, i.e. the
singularities of the unit vector $\vec n(u)$. The charges of them are
topologically quantized.

\section{From GBC theorem to Morse theory}

In this section we will study the relation between Euler characteristic $%
\chi (M)$ and indices of the critical points in Morse theory via the
topological structure. We will show that the formula of $\chi (M)$ in Morse
theory is only a corollary of the GBC theorem.

Let $f$ be an arbitrary function on $M$. A critical point of $f$ is a point $%
p\in M$ at which $df$ vanishes
\begin{equation}  \label{us1}
df|_p=\partial _afdu^a|_p=0,
\end{equation}
and at such point the Hessian $H_f(p)$ is well defined quadratic form on $%
T_pM,$ the tangent space to $M$ at $p.$ In local coordinates $\{u^a\}$
centered at $p,$ the matrix of $H_f(p)$ relative to the base $\partial _a$
at $p$ is then given by
\[
\{H_f(p)\}_{ab}=\frac{\partial ^2f}{\partial u^a\partial u^b},
\]
which is called the Hessian matrix. If there exist $l$ critical points on $f$%
, we denote them by $p_i(i=1,2,...,l).$ The index of $p_i$ is the number of
negative eigenvalues of $detH_f(p_i)$ and it will be denoted by $\lambda
_i(f).$

Now let the smooth vector field $\vec \phi $ be a gradient field\cite
{Milnor2} of the function $f$ on $M$ as
\begin{equation}  \label{fe1}
\phi ^A=e^{Aa}\partial _af,
\end{equation}
which means that the critical points of $f$ are just the zero point of $\vec %
\phi .$ From Eq. (\ref{fe1}), we get
\begin{equation}  \label{fe2}
\partial _b\phi ^A|_p=e^{Aa}\partial _a\partial _bf|_p.
\end{equation}
In terms of the above equation and the formula
\[
\varepsilon _{A_1A_2...A_n}e^{A_1a_1}e^{A_2a_2}...e^{A_na_n}=\varepsilon
^{a_1a_2...a_n}\frac 1{\sqrt{g}},
\]
one can find
\[
D(\frac \phi u)|_p=\frac 1gdetH_f(p).
\]
Therefore the GBC form and Euler characteristic $\chi (M)$ can be
represented in terms of the Hopf indices $\beta _i$ and the Hessian $%
H_f(p_i) $%
\[
\vec n^{*}d\Omega =\sum_{i=1}^l\beta _i\delta (\vec u-\vec p_i)\frac{%
detH_f(p_i)}{|detH_f(p_i)|}\sqrt{g}d^nu,
\]
and
\begin{equation}  \label{he1}
\chi (M)=\sum_{i=1}^l\beta _i\frac{detH_f(p_i)}{|detH_f(p_i)|}.
\end{equation}

In the Morse theory \cite{Milnor2,Nash} it is well-known that Morse function
$f$ has only the non-degenerate critical points $p$ at which $f$ satisfies
\[
detH_pf\neq 0.
\]
At the neighborhood of any critical point $p_i$, Morse function $f$ can take
following form
\begin{equation}  \label{mo1}
f=f(p_i)-(u^1)^2-...-(u^{\lambda _i})^2+...+(u^n)^2,
\end{equation}
where $\lambda _i=0,1,...,N.$ Substituting (\ref{mo1}) into (\ref{he1}), one
can get the generalized expression of $\chi (M)$ in the Morse theory
\begin{equation}  \label{la1}
\chi (M)=\sum_{i=1}^l\beta _i(-1)^{\lambda _i},
\end{equation}
when $f$ is taken as a Morse function. In the case of $\beta _i=1,$ one can
get the common Morse theory formula of $\chi (M)$%
\begin{equation}  \label{non-gen}
\chi (M)=\sum_{i=1}^l(-1)^{\lambda _i}.
\end{equation}
Since the meaning of Hopf index $\beta _i$ is that when the point $\vec u$
covers the neighborhood of the zero $z_i$ on $U_i$ once, the vector field $%
\vec \phi $ covers the corresponding region $\beta _i$ times, we can think
that the Hopf index corresponding to some physical degeneracy. The formula (%
\ref{non-gen}) is only the special case of non-degeneracy.

\section{Acknowledgments}

This work was supported by the National Natural Science Foundation of China
and the Doctoral Foundation of People's Republic of China.

\section{Appendix: The coefficient in Eq. (\ref{nwe1})}

Here we give a proof of the Eq. (\ref{nwe1}). From Eqs. (\ref{h3}) and (\ref
{ds0}) one can obtain
\begin{equation}
H^A=k(u)dn^A.  \label{kdn}
\end{equation}
In terms of the Eqs. (\ref{nor1}), (\ref{h2}) and (\ref{per1}) one can get
\[
dN^\mu =-B_a^\mu e^{Aa}H^A.
\]
From Eq. (\ref{kdn}) we can find the relationship
\begin{equation}
dN^\mu =-k(u)B_a^\mu e^{Aa}dn^A,  \label{na1}
\end{equation}
where $n^A$ are the unit tangent vectors which satisfy
\begin{equation}
n^An^A=1.  \label{nn1}
\end{equation}
It is well-known that the area element of $S^n$ is
\[
ds=\frac 1{n!}\varepsilon _{\lambda \mu _1\mu _2...\mu _n}N^\lambda dN^{\mu
_1}\wedge dN^{\mu _2}\wedge ...\wedge dN^{\mu _n}.
\]
In order to discuss the coefficient we introduce the following $n$-form
\[
I=\frac 2{A(S^n)n!}\varepsilon _{\lambda \mu _1\mu _2...\mu _n}N^\lambda
dN^{\mu _1}\wedge dN^{\mu _2}\wedge ...\wedge dN^{\mu _n},
\]
in terms of Eq. (\ref{na1}) which can be expressed as
\begin{equation}
I=\frac 2{A(S^n)n!}k^n(u)\varepsilon _{A_1A_2...A_n}dn^{A_1}\wedge
dn^{A_2}\wedge ...\wedge dn^{A_n}.  \label{na2}
\end{equation}
We split $S^n$ into hemispheres $S^{\pm },$ i.e. $S^n=S^{+}+S^{-}$, and $%
\partial S^{+}=S^{n-1},$ we have
\begin{eqnarray}
\int_{S^{+}}I=&&\frac 2{A(S^n)}\int_{S^{+}}ds  \nonumber \\
&&=\frac 2{A(S^n)}\frac 12A(S^n)  \nonumber \\
&&=1.  \label{a19}
\end{eqnarray}
For the unit vector $n^A$, from Eq. (\ref{nn1}) we have
\[
n^Adn^A=0,
\]
which can be considered as the linear homogeneous system of equations about $%
n^A.$ In case that $n^A$ has the nontrivial solution, i.e. $n^A$ has no
singularity in the Riemannian manifold $M,$ there is
\[
det(\partial _an^A)=0,
\]
or
\[
\varepsilon ^{a_1a_2...a_n}\varepsilon _{A_1A_2...A_n}\partial
_{a_1}n^{A_1}\partial _{a_2}n^{A_2}...\partial _{a_n}n^{A_n}=0.
\]
Thus we have $I=0,$ i.e. the coefficient $k$ in the Eq. (\ref{kdn}) can be
treated arbitrarily in the non-singularities, so we only need to consider
the $k$ in the singularity. Suppose that there are $l$ isolated
singularities in the $M,$ the $i$th singularity is
\[
u=z_i,\;\;\;i=1,2,...,l.
\]
We know that there is at least a singularity when we enclose the sphere
using a open face, and set this singularity is $z_i$. With Eq. (\ref{na2})
and Stokes' theorem we have
\begin{eqnarray}
\int_{S^{+}}I&&=\frac 2{A(S^n)}\int_{S^{+}}\frac 1{n!}k^n\varepsilon
_{A_1A_2...A_n}dn^{A_1}\wedge dn^{A_2}\wedge ...\wedge dn^{A_n}  \nonumber \\
&&=\frac 2{A(S^n)}\frac{k^n(z_i)}{n!}\int_{S^{n-1}}\varepsilon
_{A_1A_2...A_n}n^{A_1}dn^{A_2}\wedge dn^{A_3}\wedge ...\wedge dn^{A_n}
\nonumber \\
&&=\frac 2{A(S^n)}\frac{k^n(z_i)}{n!}(n-1)!A(S^{n-1}).  \label{sin1}
\end{eqnarray}
Considering the Eq. (\ref{a19}) one can find
\begin{equation}
k^n(z_i)=\frac{A(S^n)n!}{2A(S^{n-1})(n-1)!}=\frac{n!!}{(n-1)!!},  \label{k1}
\end{equation}
which shows the relationship of the Gauss mappings between the $n$-dimension
and the $(n+1)$-dimension. Obviously the Eq. (\ref{sin1}) is valid in other
singularities, in the other words, we have the result (\ref{k1}) for all
singularities. Then we obtain the relation
\[
H^A=(\frac{n!!}{(n-1)!!})^{1/n}dn^A,
\]
this is just the Eq. (\ref{nwe1}).

Here we give a proof of the Eq. (\ref{nwe1}). In order to discussing the
coefficient we introduce the following n-form
\begin{equation}  \label{na2}
I=\frac 2{A(S^n)n!}k^n(u)\varepsilon _{A_1A_2...A_n}dn^{A_1}\wedge
dn^{A_2}\wedge ...\wedge dn^{A_n}.
\end{equation}
For the unit vector $n^A$, from Eq. (\ref{cin}) we have
\[
n^Adn^A=0,
\]
which can be considered as the linear homogeneous system of equations about $%
n^A.$ In case that $n^A$ has the nontrivial solution, i.e. $n^A$ has no
singularity in the Riemannian manifold $M,$ there is
\[
det(\partial _an^A)=0,
\]
or
\[
\varepsilon ^{a_1a_2...a_n}\varepsilon _{A_1A_2...A_n}\partial
_{a_1}n^{A_1}\partial _{a_2}n^{A_2}...\partial _{a_n}n^{A_n}=0.
\]
Thus we have $I=0,$ i.e. the coefficient $k$ can be treated arbitrarily in
the non-singularities, so we only need to consider the $k$ in the
singularity for the $n$-form $I$. Considering the property of manifold, each
closed neighborhood $M_i$ of singularity can always be immersed in a ($n+1$%
)-dimensional Euclidean space $R^{n+1}$\cite{Eisenhart}. In this Euclidean
space $R^{n+1}$ there exist a vector $N^\mu $%
\[
N^\mu =\frac 1{n!}\frac 1{\sqrt{g}}\epsilon ^{\mu \mu _1\mu _2...\mu
_n}\epsilon ^{a_1a_2...a_n}B_{a_1}^{\mu _1}B_{a_2}^{\mu _2}...B_{a_n}^{\mu
_n},\;\;\;\;\mu =1,2,...,n+1
\]
which is normal to $M_i,$ i.e. $N^\mu B_a^\mu =0$. Similarly we have
\[
H^A=-B_a^\mu e^{Aa}dN^\mu .
\]
From Eqs. (\ref{h3}) one can obtain
\begin{equation}  \label{kdn}
H^A=kdn^A.
\end{equation}
then,
\begin{equation}  \label{na1}
dn^A=-k^{-1}dN^\mu B_a^\mu e^{Aa}.
\end{equation}
In terms of Eq. (\ref{na1}) the $n$-form $I$ can be expressed as
\[
I=\frac 2{A(S^n)n!}\varepsilon _{\lambda \mu _1\mu _2...\mu _n}N^\lambda
dN^{\mu _1}\wedge dN^{\mu _2}\wedge ...\wedge dN^{\mu _n}.
\]
Let $S^n=S^{+}+S^{-}$, and $\partial S^{+}=S^{n-1},$ we have
\begin{equation}  \label{a19}
\begin{array}{c}
\int_{S^{+}}I= \frac 2{A(S^n)}\int_{S^{+}}ds \\
= \frac 2{A(S^n)}\frac 12A(S^n) \\
=1.
\end{array}
\end{equation}
where $ds$ is the area element of $S^n$%
\[
ds=\frac 1{n!}\varepsilon _{\lambda \mu _1\mu _2...\mu _n}N^\lambda dN^{\mu
_1}\wedge dN^{\mu _2}\wedge ...\wedge dN^{\mu _n}.
\]
Suppose that there are $l$ isolated singularities in the $M,$ the $i$th
singularity is
\[
u=z_i,\;\;\;i=1,2,...,l.
\]
We know that there is at least a singularity when we enclose the sphere
using a open face, and set this singularity is $z_i$. With Eq. (\ref{na2})
and Stokes' theorem we have
\begin{equation}  \label{sin1}
\begin{array}{c}
\int_{S^{+}}I= \frac 2{A(S^n)}\int_{S^{+}}\frac 1{n!}k^n\epsilon
_{A_1A_2...A_n}dn^{A_1}\wedge dn^{A_2}\wedge ...\wedge dn^{A_n} \\
= \frac 2{A(S^n)}\frac{k^n}{n!}\int_{S^{n-1}}\epsilon
_{A_1A_2...A_n}n^{A_1}dn^{A_2}\wedge dn^{A_3}\wedge ...\wedge dn^{A_n} \\
=\frac 2{A(S^n)}\frac{k^n}{n!}(n-1)!A(S^{n-1}).
\end{array}
\end{equation}
Considering the Eq. (\ref{a19}) one can find
\begin{equation}  \label{k1}
k^n=\frac{A(S^n)n!}{2A(S^{n-1})(n-1)!}=\frac{n!!}{(n-1)!!},
\end{equation}
which shows the relationship of the Gauss mappings between the $n$-dimension
and the $(n+1)$-dimension. Obviously the Eq. (\ref{sin1}) is valid in other
singularities, in the other words, we have the result (\ref{k1}) for all
singularities. Then for the $n$-form $\Lambda $ we obtain the relation
\[
H^A=(\frac{n!!}{(n-1)!!})^{1/n}dn^A,
\]
this is just the Eq. (\ref{nwe1}).

\end{document}